\documentstyle[11pt]{article}
\begin{document}
{\bf RECOVERING THE INTERNAL DYNAMICS AND THE SHAPES OF GALAXY CLUSTERS:
VIRGO CLUSTER}
\vspace{0.2in}

V.G.Gurzadyan$^1$, S.Rauzy$^2$ and R.Triay$^3$

\vspace{0.2in}

1. ICRA, Dipartimento di Fisica, Universita di Roma La Sapienza,
Rome, Italy; Yerevan Physics Institute and Garny Space Astronomy
Institute, Armenia.

2. Department of Physics and Astronomy, University of Glasgow,
Glasgow, G12 8QQ, UK.

3. Universit\'e de Provence and 
Centre de Physique Th\'eorique, 
C.N.R.S. Luminy, Case 907, F-13288 Marseille
Cedex 9, France.
\vspace{0.2in}

We describe a method for recovering of
the substructure, internal dynamics and geometrical shapes
of clusters of galaxies. 
Applying the method to the Virgo cluster, we
first,  reveal the substructure of the central 4 arc degree field of the Virgo cluster
by means of S-tree technique. The existence of three main subgroups
of galaxies is revealed and their dynamical characteristics are estimated.
Then, using the previously suggested technique (Ref.\cite{GR97}),
the bulk flow velocities of the
subgroups are evaluated based on the distribution of the redshifts of the
galaxies. The results enable us also to obtain
a secure indication of the elongation of the Virgo cluster and its
positional inclination.
\vspace{0.2in}


\section{Introduction}
\label{Introduction}

The present paper aims to find out a method enabling
the study of the subgroupings and their bulk flows, as well as
the shapes of the clusters of galaxies, based on rather general
assumptions. The existence of subgroups of clusters in general is a known fact,
revealed by different methods for various clusters  (see e.g.$^{\cite{Gi}}$). 
The study of a sample of Abell clusters from ESO Key 
program (ENACS) performed by S-tree method enabled to conclude the existence
of subgroups -  {\it galaxy associations},  dynamical entities with
remarkable properties as a possible common feature of clusters of 
galaxies$^{\cite{GurM},\cite{GurM97},\cite{GurM01}}$.  

We use the method to
analyze the substructure and internal dynamics of the
central region of Virgo cluster.
Though the Virgo cluster because of its close location is one of the 
best studied clusters, the precise measurements of distance have been 
performed for relatively few galaxies which have been used for the 
accurate determination of the Hubble constant. These observational
studies at present are being intensely continued (see e.g. Ref.\cite{Fed},\cite{West}), 
so that the most accurate study
of the internal dynamics of the cluster is especially desirable, since
each galaxy can participate a bulk flow motion, apart from the Hubble
expansion.

To study the internal dynamics one has first to analyze the hierarchical
structure of the cluster.
By this first step we reveal the 
substructure of the Virgo cluster
core within the $\pm$2 arc degree 
field centered on M87 using the S-tree method. 
On this point the present study is the generalization of the 
previous study$^{\cite{Petr}}$, 
where the Virgo cluster within central 1 arc degree field has been studied.
For the second step - the reconstruction of the bulk regular flow of the 
subgroups,
we use the method developed in Ref.\cite{GR97} based
on the data on the radial velocities of the members of the cluster. This
method has been already applied to reveal the bulk motions 
within the Local Group
of galaxies, and hence to obtain the apex and the 3D velocity vector of  the motion
of the Local Group with respect to the
Cosmic Microwave Background frame$^{\cite{GR98}}$.

For the Virgo cluster
our analysis indicates the existence of 3 main subgroups of galaxies.
Then, we estimate the transversal components of regular motions of the 
subgroups with respect to each other.
Among the main results of this study we have to mention the
unambiguous indication of the elongated nature of the Virgo cluster, and
combined $^{\cite{Petr}}$with the rest kinematic parameters of the system we could
even get constraints on the inclination angle and the degree of elongation.

The ongoing studies on better estimation of the distances of individual 
galaxies, particularly using the Cepheids, Tully-Fisher
methods (e.g. $^{\cite{Yo},\cite{GF},\cite{Bo}}$, the  method of Surface Brightness Fluctuations for Virgo's brightest galaxies$^{\cite{West}}$, 
and other methods will enable from the results obtained here to determine more precisely the 
3D bulk flow of each subgroups, and thus the mean motion of the
Virgo cluster itself. Essential complementary information on the parameters
of the clusters and groups of galaxies can be provided by the X-ray data$^{\cite{Wu}}$.

\section{Virgo cluster substructures}

To study the substructures of the Virgo cluster core we used the data from
the CfA redshift catalogue$^{\cite{Hu}}$. We dealt with 
the four arc degree central field within:
N4316, RA= 12h 20m 12s  and N4584, RA= 12h 35m 46.8s  
with the Virgo center coordinates RA=12h 27m 50.4s, DEC=12d 55m 55.2s.
The sample includes 146 galaxies with available redshifts and V magnitudes.

For the description of S-tree method one can refer not only to the 
original works$^{\cite{Gur94a},\cite{Gur94b}}$, but also to the recent 
applications of this method to various clusters$^{\cite{GR97},\cite{GurM97}}$,
where it is discussed  in
details. Therefore here we only stress that this method is revealing 
the degree of mutual interaction of members of the system using 
self-consistently the information both on the 2D coordinates and redshifts 
of individual galaxies of the cluster as well as on their magnitudes.

The results of our study of the Virgo cluster core showed the 
existence of three main subgroups of galaxies containing 54, 17 
and 37 galaxies, 
correspondingly. The lists of those galaxies along with their coordinates
and redshifts are given in Tables 1, 2 and 3.

\begin{table}
\caption{VIRGO subgroup I: 54 galaxies.}
\begin{center}
\begin{tabular}{lllr|lllr}
\hline
Name & $l$ & $b$ & $z_{\rm CMB}$ &
Name & $l$ & $b$ & $z_{\rm CMB}$ \\
\hline
N4316        & 280.722 & 70.963 & 1599  & 
N4371        & 279.666 & 73.369 & 1273  \\ 
N4374        & 278.183 & 74.475 & 1363  & 
N4377        & 275.310 & 76.206 & 1700  \\ 
N4379        & 273.742 & 76.967 & 1394  & 
N4380        & 281.900 & 71.820 & 1298  \\ 
N4390        & 281.819 & 72.270 & 1452  & 
I3328        & 282.325 & 71.897 & 1337  \\ 
I3331        & 280.458 & 73.575 & 1608  & 
A1223+1308   & 279.110 & 74.545 & 1594  \\ 
N4411        & 283.890 & 70.818 & 1616  & 
I3344        & 278.454 & 75.247 & 1702  \\ 
N4411        & 284.091 & 70.856 & 1605  & 
N4417        & 283.452 & 71.520 & 1178  \\ 
I3356        & 281.390 & 73.389 & 1433  & 
I3371        & 282.534 & 72.783 & 1260  \\ 
N4429        & 282.361 & 73.012 & 1463  & 
N4431        & 280.995 & 74.132 & 1243  \\ 
I3374        & 283.584 & 71.970 & 1202  & 
N4436        & 281.176 & 74.175 & 1454  \\ 
A1225+900    & 285.316 & 70.800 & 1442  & 
N4451        & 285.083 & 71.334 & 1195  \\ 
A1226+94     & 285.139 & 71.512 & 1374  & 
N4459        & 280.123 & 75.842 & 1540  \\ 
I3413        & 283.586 & 73.469 & 1735  & 
N4477        & 281.538 & 75.610 & 1680  \\ 
N4478        & 283.377 & 74.392 & 1698  & 
N4479        & 281.855 & 75.576 & 1183  \\ 
N4483        & 286.771 & 71.223 & 1215  & 
A1228+1219   & 284.137 & 74.159 & 1578  \\ 
N4486        & 283.761 & 74.489 & 1620  & 
N4497        & 285.185 & 73.806 & 1427  \\ 
I3457        & 284.374 & 74.818 & 1796  & 
N4503        & 286.081 & 73.413 & 1688  \\ 
I3468        & 287.006 & 72.516 & 1703  & 
I3468        & 285.632 & 74.122 & 1438  \\ 
A1230+926    & 288.084 & 71.497 & 1563  & 
N4515        & 280.529 & 78.319 & 1258  \\ 
N4516        & 283.225 & 76.739 & 1280  & 
I3487        & 288.443 & 71.748 & 1489  \\ 
N4519        & 289.177 & 71.049 & 1562  & 
I3499        & 287.655 & 73.337 & 1555  \\ 
N4528        & 287.628 & 73.675 & 1702  & 
I3510        & 287.986 & 73.446 & 1703  \\ 
I3518        & 289.268 & 72.045 & 1771  & 
I3520        & 285.897 & 75.816 & 1416  \\ 
N4540        & 283.618 & 77.801 & 1605  & 
A1232+927    & 289.947 & 71.644 & 1617  \\ 
N4551        & 288.150 & 74.681 & 1523  & 
N4564        & 289.541 & 73.920 & 1492  \\ 
I3583        & 288.272 & 75.709 & 1448  & 
I3602        & 291.863 & 72.660 & 1607  \\ 
\end{tabular}
\end{center}
\label{table1}
\end{table}

\begin{table}
\caption{VIRGO subgroup II: 17 galaxies.
}
\begin{center}
\begin{tabular}{lllr|lllr}
\hline
Name & $l$ & $b$ & $z_{\rm CMB}$ &
Name & $l$ & $b$ & $z_{\rm CMB}$ \\
\hline
N4321        & 271.133 & 76.899 & 1884   &
N4330        & 278.752 & 72.906 & 1885   \\
N4383        & 272.099 & 77.758 & 2015   &
N4405        & 273.413 & 77.585 & 2063   \\
I3349        & 280.227 & 74.227 & 1801   &
N4421        & 275.764 & 77.021 & 1925   \\
I3369        & 274.990 & 77.549 & 2048   &
I3392        & 278.247 & 76.760 & 2001   \\
I796         & 276.345 & 78.126 & 1919   &
N4474        & 280.797 & 76.005 & 1949   \\
N4486B       & 283.392 & 74.563 & 1914   &
I3457        & 284.374 & 74.818 & 1796   \\
I3470        & 286.240 & 73.508 & 1829   &
I3501        & 285.437 & 75.594 & 1932   \\
I3586        & 289.098 & 75.005 & 1892   &
N4578        & 291.682 & 72.125 & 2613   \\
N4579        & 290.374 & 74.361 & 1845   &
             &         &        &        \\
\end{tabular}
\end{center}
\label{table2}
\end{table}

\begin{table}
\centering
\caption{VIRGO subgroup III: 37 galaxies.
}
\begin{center}
\begin{tabular}{lllr|lllr}
\hline
Name & $l$ & $b$ & $z_{\rm CMB}$ &
Name & $l$ & $b$ & $z_{\rm CMB}$ \\
\hline
I3239        & 278.239 & 73.229 &  979   &
N4328        & 271.511 & 76.943 &  841   \\
N4387        & 278.828 & 74.465 &  913   &
A1223+1513   & 275.513 & 76.439 &  830   \\
N4402        & 278.760 & 74.783 &  565  &
I3363        & 280.318 & 74.349 & 1120   \\
N4424        & 283.865 & 71.389 &  767   &
A1224+936    & 284.169 & 71.332 & 1047   \\
N4435        & 280.148 & 74.889 & 1101   &
N4440        & 281.363 & 74.170 & 1068   \\
N4442        & 284.140 & 71.819 &  849   &
N4445        & 284.625 & 71.482 &  635   \\
I3381        & 282.244 & 73.722 &  967   &
I3393        & 281.313 & 74.817 &  794   \\
N4452        & 282.696 & 73.729 &  553   &
N4458        & 281.078 & 75.151 & 1011   \\
I3412        & 284.978 & 72.086 & 1098   &
A1226+1243   & 282.451 & 74.438 &  866   \\
N4469        & 286.087 & 70.905 &  833   &
N4486A       & 283.984 & 74.386 &  778   \\
N4491        & 284.808 & 73.636 &  826   &
I3459        & 284.941 & 74.359 &  606   \\
I3466        & 285.451 & 74.028 & 1114   &
N4506        & 283.826 & 75.578 & 1006   \\
I798         & 281.397 & 77.480 &  734   &
N4523        & 283.106 & 77.362 &  582   \\
I3517        & 289.588 & 71.588 &  764   &
I3522        & 284.036 & 77.476 &  980   \\
N4548        & 285.669 & 76.830 &  871   &
I3540        & 287.350 & 75.326 & 1057   \\
N4550        & 288.087 & 74.635 &  706   &
A1233+1239   & 288.036 & 74.793 &  608   \\
N4552        & 287.914 & 74.967 &  647   &
A1233+1408   & 286.827 & 76.240 & 1078   \\
I3578        & 289.978 & 73.592 & 1021   &
N4571        & 287.486 & 76.659 &  663   \\
A1235+1508   & 288.207 & 77.363 & 1055   &
             &         &        &        \\
\end{tabular}
\end{center}
\label{table3}
\end{table}

While comparing the results of this analysis with those of the 1 degree 
field$^{\cite{Petr}}$
it is interesting to notice, that we see the strong
correspondence with the 1 degree field membership results.
This fact is remarkable since it shows the robust character of the results
of subgrouping derived by S-tree method even when limited areas of the 
clusters are studied. Relative stability of the
subgrouping at various scales follows also from physical considerations.

As seen from Table 4 below, though the mean redshifts
of the subgroups are rather  different, their redshift dispersions are
quite close to each other. While the
study of the regular motions of the subgroups will be
done in next sections, already the remarkable difference of line-of-sight 
velocities is already indicating their possible significant
radial bulk velocity 
with respect to each other.

\section{Bulk flow reconstruction}

The basic idea of the reconstruction procedure of the 3D motions within an
N-body system from data on 1D line-of-sight velocities relies on
the existence of correlation
between the velocities of different members of an interacting system. From this
point of view our scheme of 3D velocity reconstruction and the S-tree method have identical physical
background. For the case
of stellar systems (large $N$) the problem of reconstruction of 3D velocity distribution
of stars has been solved decades ago by Ambartsumian$^{\cite{Amb36}}$, 
without any a priori assumption on the form of the distribution
function. The only natural assumption was its translation invariance,
i.e. the phase space distribution of the system can be split as
\begin{equation}\label{dPphasespacetext}
dP=\Phi_{\rm S}(x_1,x_2,x_3)dx_1dx_2dx_3
\times \Phi_{\rm K}(v_1,v_2,v_3)dv_1dv_2dv_3,
\end{equation}
where $\Phi_{\rm S}(x_1,x_2,x_3)$ is the 3D spatial distribution of
objects and
$\Phi_{\rm K}(v_1,v_2,v_3)$  is the 3D velocity distribution function.
The efficiency of such reconstruction procedure is however
closely related to the number of objects considered.
For the case of galaxy clusters (small $N$), 
it turns out that Ambartsumian's method cannot be applied directly.
However some quantities of interest, for example the
bulk flow of individual clusters, can be extracted from
the data by assuming reasonable form of
the velocity distribution function $\Phi_{\rm K}$. Herein 
we assume that the velocities
distribution of galaxies inside the cluster
can be  described by gaussian random 
isotropic components of velocity dispersion
$\sigma_v$ plus the 3D mean peculiar
velocity of the cluster, i.e.
\begin{equation}\label{dPKtext}
\Phi_{\rm K}(v_1,v_2,v_3)=g(v_1;V_1,\sigma_v)\,
g(v_2;V_r,\sigma_v) \,
g(v_3;V_3,\sigma_v),
\end{equation}
where $V_r$ is the radial component of the cluster bulk flow
and $V_1$ and $V_3$ are its components perpendicular to
the line-of-sight.
Note however that this hypothesis does not imply any 
severe limitations in our
problem. Moreover it is indeed justified by the fact that due to exponential 
instability of gravitating systems and their mixing properties, the 
correlations in physical parameters of particles have to split. This effect
can be followed in numerical experiments.

Another limitation comes from the fact that galaxies
inside a cluster
take part in the Hubble flow, implying then that a fraction  of
the observed redshift is due to the spatial elongation of the
cluster along the line-of-sight. A common practice is
to assume that this effect is negligible, i.e. all the galaxies
lie at the same distance. 
In Ref.\cite{GR97} we used a more physical
assumption allowing 
a 3D spatial extension of the cluster with  
gaussian isotropic distribution around the cluster center, i.e.
\begin{equation}\label{dPStext}
\Phi_{\rm S}(x_1,x_2,x_3)=g(x_1;0,\sigma_S)\,
g(x_2;0,\sigma_S) \,
g(x_3;0,\sigma_S),
\end{equation}
where $\sigma_S$ is the spatial dispersion of the cluster.
Under this working assumption it is shown in Appendix A3 that
the observed
probability density reads in terms of the redshift $z$ and the angular
position with respect to the center of the cluster
${\bf \theta}=(\theta_1,\theta_3)$ of the galaxies as
 	\begin{equation}\label{dPobstext} 
dP_{\rm obs} =g(z- \langle z \rangle; V_1\,\theta_1 
+V_3 \,\theta_3 ,\sigma_{\rm obs})
  \times g(\theta_1; 0 ,\sigma_{\theta_1})
\, g(\theta_3; 0,\sigma_{\theta_3})\,d\theta_3 d\theta_1 dz
	\end{equation}
where $\sigma_{\rm obs}^2=\sigma_{v}^2+\sigma_{S}^2$,
$\langle z \rangle$ is the
mean redshift of the cluster and $\sigma_{\theta_1}$ and 
$\sigma_{\theta_3}$ are the angular sizes of the cluster
in the plane perpendicular to the mean line-of-sight.
It thus turns out from Eq. (\ref{dPobstext}) that the
tangential velocity of the cluster $V_{\rm tan}=(V_1,V_3)$
can be evaluated by using standard statistical technique.

\section{Application to Virgo subgroups}

For each Virgo subgroup the system of coordinates 
$(x_1,x_2,x_3)$ is defined as follows.
The $x_2$-axis is directed towards the mean angular
position of subgroup galaxies ($ \langle l \rangle$,$ \langle b \rangle$),
the $x_1$-axis is parallel to the galactic plane and the $x_3$-axis
is chosen in a way that the $(x_1,x_2,x_3)$ system forms a direct
trihedron. The $x_1$, $x_2$ and $x_3$ coordinates then are  
deduced by transforming the cartesian galactic coordinates
($z \cos l \cos b$,$z \sin l \cos b$,$z \sin b$) so that
the two trihedrons coincide. Because the velocity dispersion
$\sigma_v$ is small compared to the mean redshift 
$\langle z \rangle$ of the cluster, 
one has within the approximation of
small angles $x_1 \approx \theta_1\,\langle z \rangle $ 
and $x_3 \approx \theta_3\,\langle z \rangle $
where $\theta_1$ and $\theta_3$ are the position angles along
the $x_1$ and $x_3$-axis respectively, and  $x_2 \approx z$.

The 3D data distribution of Virgo subgroup I is given in Figure
1, where the $x_2$ coordinate has been transformed to
$x_2  = z - \langle z \rangle$ for convenience. Redshifts are
expressed in km s$^{-1}$ in the CMB frame. 
The influence of peculiar velocities is clearly seen
on $x_1x_2$ and $x_2x_3$ projections. Indeed 
the data distribution is
much more elongated along the $x_2$-axis due to
the presence of the velocity dispersion.
The two left panels of the figure reveal a correlation
between $\theta_1$ and $\theta_3$ (or equivalently
between $x_1$ and $x_3$). Such a correlation
which cannot be described by  Eq. (\ref{dPobstext})
clearly means that we have to modify our initial hypothesis, namely that
the spatial distribution of the subgroup galaxies admits
a central symmetry. 

\vspace{0.2in}

FIGURE 1.

\vspace{0.2in}

Indeed the hypothesis of a central symmetry 
is stronger than it is necessary for the problem
we are dealing with
i.e. for the tangential bulk flow estimation. 
In Appendix A3 it is shown that the sufficient condition
for evaluating $V_{\rm tan}$  is the absence of correlation
between the spatial distribution  of galaxies along the line-of-sight and
perpendicular to it, i.e. $a_{21}$ and $a_{23}$ are vanishing in Eq.
(\ref{A21A23}).
In this case, the observed distribution function takes the
form
 	\begin{equation}\label{dPobstext1} 
\begin{tabular}{ll}
$dP_{\rm obs}$ & $=g(z- \langle z \rangle; A_{21}\,\theta_1 
+A_{23} \,\theta_3 ,\sigma_{\rm obs}) $
\\ \\
 & $\, \times g(\theta_1; A_{13}\,\theta_3 ,\sigma'_{\theta_1})
\, g(\theta_3; 0,\sigma_{\theta_3})\,d\theta_3 d\theta_1 dz
$ 
 \end{tabular} 
	\end{equation}
where the tangential velocities are $V_1 \equiv A_{21}$ and 
$V_2 \equiv A_{23}$. 
Here the parameter
$A_{13}$ allows us to include into consideration  the 
correlation between $\theta_1$ and $\theta_3$,
as required by the observational data.

The estimates of tangential bulk flows of the three Virgo subgroups
are given in Table 4 using the formulae derived in Appendix A3.
Due to the relatively small number of galaxies in the subgroups the error 
boxes are large, but nevertheless are informative.
Note, that for subgroups I and II the bulk flow peculiar tangential velocities
are anomaly high. These values are significant i.e. are
not due to the sampling errors.
\begin{table}
\centering
\caption{Estimate of the parameters for the 3 Virgo subgroups.
Units: 
$ \langle l \rangle$ and
$ \langle b \rangle$ in degrees,
$\sigma_{\theta_1}$,
$\sigma_{\theta_3}$ and
$\sigma'_{\theta_1}$ in radians and
$ \langle z \rangle$,
$\sigma_{\rm obs}$,
$A_{21}$ and 
$A_{23}$ in km s$^{-1}$. 
}
\begin{tabular}{l|c|c|c}
 & Virgo I & Virgo II & Virgo III \\
\hline
$N_{\rm gal}$     & $54$  & $17$ &  $37$  \\
\hline
$ \langle z \rangle$       & $1497.91$ & $1958.22$ & $862.96$   \\
$ \langle l \rangle$       & $284.162$ & $281.446$ & $283.635$   \\
$ \langle b \rangle$       & $73.658$ & $75.637$ & $74.522$   \\
\hline
$\sigma_{\theta_1}$ & $0.0182$ & $0.0302$ & $0.0176$ \\
$\sigma_{\theta_3}$ & $0.0326$ & $0.0338$ & $0.0331$ \\
$\sigma'_{\theta_1}$ & $0.0165$ & $0.0183$ & $0.0173$ \\
\hline
$\rho_{13}$ & $-0.172$ & $-0.631$ & $-0.033$\\
$A_{13}$ & $-0.23 \pm 0.07$ &  $-0.71 \pm 0.14$ & $-0.10 \pm 0.09$ \\
\hline
$\sigma_{\rm obs}$ & $167.22$ & $171.26$  & $178.95$ \\
\hline
$A_{21}$ & $2485 \pm 1375$ & $2684 \pm 2264$ & $-489 \pm 1703$\\
$A_{23}$ & $938 \pm 767$ & $476 \pm 2023$ & $-109 \pm 904$ \\

\end{tabular}
\label{table4}
\end{table}

\section{The spatial structure of Virgo subgroups}

The statistically significant correlation between
$\theta_1$ and $\theta_3$ for Virgo subgroups I and II 
advocates  against 
the hypothesis 
of central symmetry for the spatial distribution of their galaxies.
This means that these systems represent spatially elongated configurations.
Their elongation can be described by the following
ellipsoidal 3D spatial distribution function
\begin{equation}\label{dPSy}
\Phi_{\rm S}(y_1,y_2,y_3)=g(y_1;0,\sigma_1)\,
g(y_2;0,\sigma_2) \,
g(y_3;0,\sigma_3)
\end{equation}
where at least one of the dispersions, say $\sigma_1$, differs
from the two others. In this case, the orientation of the
proper system of coordinates $(y_1,y_2,y_3)$ of the ellipsoidal
distribution has to be defined with respect to the
frame of analysis, namely
$(x_1,x_2,x_3)$. This is done by introducing
3 rotation angles $\alpha$, $\beta$
and $\gamma$, where
the angles $\alpha$ and $\beta$ are respectively the
longitude and the latitude of the $y_1$-axis in the 
$(x_1,x_2,x_3)$ frame and $\gamma$ is the
angle between 
the intersection
of the planes $x_1x_2$ and $y_2y_3$ and
the $y_2$-axis. 
As it is shown in Appendix A2  the spatial probability density
in the coordinate frame 
$(x_1,x_2,x_3)$ has to form 
 	\begin{equation}\label{dPSxtext} 
dP_{\rm S}=g(x_3;0,\sigma_{\rm I}) \, g(x_1;a_{13}\,x_3,\sigma_{\rm II})
 \times g(x_2; a_{21}\,x_1+a_{23} \,x_3 ,\sigma'_{\rm III})\,dx_1 x_2 x_3 
	\end{equation}
where the parameters $a_{13}$, $a_{21}$, $a_{23}$,
$\sigma_{\rm I}$, $\sigma_{\rm II}$ and $\sigma'_{\rm III}$
are linked to the proper spatial dispersions
$\sigma_1$, $\sigma_2$ and $\sigma_3$ and the orientation
angles $\alpha$, $\beta$ and $\gamma$ by the formulae given in that appendix.
The case $a_{21}=0$ and $a_{23}=0$ arises if the ellipsoid
is symmetrical with respect to the plane perpendicular to
the line-of-sight ($\alpha=0$ and $\gamma=0$).

A it is shown in the appendix A3,  
the observed distribution function has the same form as
in Eq. (\ref{dPobstext1}), but then the parameters
$A_{21}$ and $A_{23}$ are
\begin{equation}\label{A21A23text}
A_{21} = V_1 + a_{21}\,H_0 \,D
\,\,\,\,\,\,\,\,\,\, ;
\,\,\,\,\,\,\,\,\,\,
A_{23} = V_3 + a_{23}\,H_0 \,D
\end{equation}
where $H_0$ is  the Hubble's constant and $D$ is
the distance of the considered subgroup.
These expressions show us that in general it is not possible
to disentangle between the presence of tangential subgroup bulk flow
and the effect of inclination along the line-of-sight of
the ellipsoidal distribution of subgroup galaxies.
In particular, it is very likely that the
large values of $A_{21}$ and $A_{23}$ obtained for the
Virgo I and II subgroups are partly due to this inclination effect.

Finally, let us consider the
spatial structure of Virgo su
bgroups I and II based on the results given in
Table 4. The initial model
contains $10$ free parameters,
namely  the 3 bulk flow components $V_r$, $V_1$ and
$V_3$, the velocity dispersion $\sigma_v$, 
the 3 proper spatial dispersions
$\sigma_1$, $\sigma_2$ and $\sigma_3$ and the 3 orientation
angles $\alpha$, $\beta$ and $\gamma$.  Our approach enables us
to estimate the $6$ independent parameters
$A_{13}$, $A_{23}$, $A_{21}$,
$\sigma'_{\theta_1}$, $\sigma_{\theta_3}$
and $\sigma_{\rm obs}$.
Assuming that the subgroup is at rest in the CMB frame,
i.e.  $V_r=0$, $V_1=0$ and $V_3=0$,  
one can deduce for a given value of  the velocity dispersion $\sigma_v$ 
the characteristics of the spatial structure of the subgroup.

We have performed this experiment for Virgo subgroups I and II
based on the data in Table 4. Figures 2 and 3
show respectively the variation of parameters
$\sigma_1$, $\sigma_2$, $\sigma_3$,
$\alpha$, $\beta$ and $\gamma$ as a function of 
$\sigma'_{\rm III}=\sqrt{\sigma_{\rm obs}^2-\sigma_v^2}$
for the subgroups I and II. Since the problem does not admit
a general analytical solution, these curves have been 
obtained numerically. 

The results show that both subgroups reveal an elongated structure with
significant difference between 
$\sigma_1$, $\sigma_2$ and $\sigma_3$, 
whatever is the contribution of the
true velocity dispersion $\sigma_v$ in the observed redshift
dispersion  $\sigma_{\rm obs}$. 

\vspace{0.2in}

FIGURES 2 and 3.

\section{Conclusion}

Thus  the technique we had developed for 3D velocity
reconstruction of galaxy configurations$^{\cite{GR98}}$ 
notwithstanding to the inevitable error boxes,
can be successfully applied to the Virgo cluster subgroups.

By a first step we have estimated the tangential bulk flow velocity of
the 3 subgroups under the assumption that the observed redshift
dispersion is not due to their 3D spatial structure.
For two of the subgroups, namely Virgo I and Virgo II, these values
were found statistically significant but anormaly high
($2656$ km s$^{-1}$ and $2726$ km s$^{-1}$ respectively). 
This fact indicates the validity of our main working hypothesis,
i.e. that the clustered galaxies exhibit an elongated spatial distribution
along the line-of-sight. 

We have shown that the knowledge of galaxies redshifts alone still does not
permit to disentangle between the presence of tangential bulk flow and the 
effect of inclination of an elongated 3D spatial structure.
Nevertheless, while both effects are certainly contributing to our derived
statistics, we can assert unambiguously that both
subgroups are spatially elongated, with the specific characteristics of their 3D
structures depending on the kinematic properties of
the subgroups. 

\section*{Acknowledgements}
We are thankful to Fang Li Zhi for valuable comments. 
V.G. acknowledges the support by French-Armenian Jumelage.

\appendix{
\section{Useful formulae}
\label{AppendixA1}

The probability density of a random
variable $x$ given as a gaussian distribution with a  mean $x_0$ and 
dispersion $\sigma_x$ 
\begin{equation}\label{Gaussian}
dP_{\rm G}=g(x;x_0,\sigma_x) \,dx =
\frac{1}{\sqrt{2 \pi}\,\sigma_x}
\exp \left( -\frac{(x-x_0)^2}{2\sigma_x^2} \right) dx
\end{equation}
fulfills the properties
\begin{equation}\label{lambdaplus}
g(x+\lambda;x_0,\sigma_x)\,dx = g(x;x_0-\lambda,\sigma_x)\,dx
\end{equation}
\begin{equation}\label{lambdatimes}
g(\lambda\,x;\lambda\,x_0,|\lambda|\,\sigma_x)\,d(|\lambda|\,x) =
g(x;x_0,\sigma_x)\,dx, 
\end{equation}
where $\lambda$ is a real number. The following
relation also holds         
\begin{equation}\label{2Gaussian}
g(x;x_1,\sigma_1)\, g(x;x_2,\sigma_2)=
g(x;x_0,\sigma_0)
\, g(x_1;x_2,\sigma),
\end{equation}
with $x_0$, $\sigma_0$ and $\sigma$ defined as
\begin{equation}\label{2Gaussiandef1}
\sigma^{2} = \sigma_1^{2}+\sigma_2^{2}
\,\,\,\,\,\,\,\,\,\, ;
\,\,\,\,\,\,\,\,\,\,
\sigma_0^{2} = \frac{\sigma_1^{2}\sigma_2^{2}}
{\sigma_1^{2}+\sigma_2^{2}}.
\end{equation}
\begin{equation}\label{2Gaussiandef2}
x_0 = \frac{\sigma_2^{2}}{\sigma_1^{2}+\sigma_2^{2}} \, x_1
+\frac{\sigma_1^{2}}
{\sigma_1^{2}+\sigma_2^{2}}\, x_2.
\end{equation}

We describe an ellipsoidal 2-dimensional distribution 
in the system of coordinates $(y_1,y_2)$
by the probability density
\begin{equation}\label{dPellipsoidal}
dP_{\rm ell}=g(y_1;0,\sigma_1)\,dy_1 \times g(y_2;0,\sigma_2)\,dy_2
\end{equation}
or using the properties (\ref{lambdaplus}), (\ref{lambdatimes}) and
(\ref{2Gaussian})  in the system of coordinates $(x_1,x_2)$ 
\begin{equation}\label{dPx1x2}
dP_{\rm ell}=g(x_1;A\,x_2,\sigma'_1)\, g(x_2;0,\sigma'_2)\,dx_1dx_2 .
\end{equation}
with $\sigma'_2$, $\sigma'_1$ and $A$ defined as follows
\begin{equation}\label{defsigmaprime}
{\sigma'_2}^2 =  \sigma_2^2 \,\cos^2 \alpha+\sigma_1^2 \,\sin^2 \alpha
\,\,\,\,\,\,\,\,\,\, ;
\,\,\,\,\,\,\,\,\,\,
\sigma'_1 = \frac{\sigma_1 \sigma_2}{\sigma'_2 }
\end{equation}
\begin{equation}\label{defA}
A = \sin \alpha \cos \alpha
\left ( \frac{ \sigma_1^{2}-\sigma_2^{2} }{ {\sigma'_2}^{2} } \right ).
\end{equation}
Here the angle $\alpha$ characterizes the rotation ${{R}}_\alpha$
transforming $(x_1,x_2)$  into $(y_1,y_2)$ coordinates, i.e.
 	\begin{equation}\label{alpharot} 
{{R}}_\alpha \,\, : \,\, \left\{
\begin{tabular}{ll}
& $y_1 = \cos \alpha \,\, x_1 + \sin \alpha \,\, x_2$ \\
& $y_2 = -\sin \alpha \,\, x_1 + \cos \alpha \,\, x_2$. 
\end{tabular} \right. 
	\end{equation}

\section{The spatial distribution}
\label{AppendixA2}

The 3-dimensional spatial
distribution of considered galaxies can be expressed
in the system of coordinates $(y_1,y_2,y_3)$ as
\begin{equation}\label{dPS}
dP_{\rm S}=g(y_1;0,\sigma_1)\,dy_1 \times g(y_2;0,\sigma_2)\,dy_2
\times g(y_3;0,\sigma_3)\,dy_3.
\end{equation}
The spatial orientation of this trihedron is herein
characterized by 3 rotation angles $\alpha$, $\beta$
and $\gamma$ with respect to the frame of analysis $(x_1,x_2,x_3)$. 
The angles $\alpha$ and $\beta$ are respectively the
longitude and the latitude of the $y_1$-axis such
that coordinate transformations read as  
 	\begin{equation}\label{Ralpha} 
 {R}_\alpha \,\, : \,\, \left\{
\begin{tabular}{ll}
& $y''_1 = \cos \alpha \,\, x_1 + \sin \alpha \,\, x_2$ \\
& $y''_2 = -\sin \alpha \,\, x_1 + \cos \alpha \,\, x_2$ \\
& $y''_3 = x_3$,
\end{tabular} \right. 
	\end{equation}
where $(y''_1,y''_2,y''_3)$ is the new system of coordinates
after a rotation ${R}_\alpha$ of angle $\alpha$ around the $x_3$-axis, 
 	\begin{equation}\label{Rbeta} 
{R}_\beta \,\, : \,\, \left\{
\begin{tabular}{ll}
& $y'_1 = \cos \beta \,\, y''_1 + \sin \beta \,\, y''_3$ \\
& $y'_2 = y''_2$ \\
& $y'_3 = -\sin \beta \,\, y''_1 + \cos \beta \,\, y''_3$
\end{tabular} \right. 
	\end{equation}
Here $(y'_1,y'_2,y'_3)$ is the system of coordinates
after rotating on an angle $\beta$ with repect the $y''_2$ axis,
and the angle $\gamma$ is defined by a rotation ${R}_\gamma$ around
$y'_1$ axis so that $y'_2$  and $y_2$ axes coincide,
i.e.
 	\begin{equation}\label{Rgamma} 
{R}_\gamma \,\, : \,\, \left\{
\begin{tabular}{ll}
& $y_1 = y'_1$ \\
& $y_2 = \cos \gamma \,\, y'_2 + \sin \gamma \,\, y'_3$ \\
& $y_3 = -\sin \gamma \,\, y'_2 + \cos \gamma \,\, y'_3$.
\end{tabular} \right. 
	\end{equation}

Our aim is to express the
density probability of Eq. (\ref{dPS}) in terms of the
coordinates of analysis, namely $(x_1,x_2,x_3)$.
Using  (\ref{dPx1x2}) and in view of 
(\ref{Rgamma}), one has
 	\begin{equation}\label{dF_1} 
\begin{tabular}{ll}
$dF_1$ & $=g(y_2;0,\sigma_2)\,dy_2 \, g(y_3;0,\sigma_3)\,dy_3$ \\
\\
 & $=g(y'_2;A\,y'_3,\sigma'_2)\, g(y'_3;0,\sigma'_3)\,dy'_2 dy'_3$, 
\end{tabular} 
	\end{equation}
with $\sigma'_3$, $\sigma'_2$ and $A$ defined as
\begin{equation}\label{sigmaprime32}
{\sigma'_3}^2 =  \sigma_3^2 \,\cos^2 \gamma+\sigma_2^2 \,\sin^2 \gamma
\,\,\,\,\,\,\,\,\,\, ;
\,\,\,\,\,\,\,\,\,\,
\sigma'_2 = \frac{\sigma_2 \sigma_3}{\sigma'_3 },
\end{equation}
\begin{equation}\label{A}
A = \sin \gamma \cos \gamma
\left ( \frac{ \sigma^{2}_2-\sigma^{2}_3 }{ {\sigma'_3}^{2} } \right ).
\end{equation}
The relation (\ref{dPx1x2}) and definitions
(\ref{Rbeta},\ref{Ralpha},\ref{Rgamma}) yield
 	\begin{equation}\label{dF_2} 
\begin{tabular}{ll}
$dF_2$ & $=g(y_1;0,\sigma_1)\,dy_1 \, g(y'_3;0,\sigma'_3)\,dy'_3$ \\
\\
 & $=g(y''_1;B\,x_3,\sigma''_1)\, g(x_3;0,\sigma''_3)\,dy''_1 dx_3$, 
\end{tabular} 
	\end{equation}
with $\sigma''_3$, $\sigma''_1$ and $B$ defined as
\begin{equation}\label{sigmaprimeprime31}
{\sigma''_3}^2 =  {\sigma'_3}^{2} \,\cos^2 \beta+\sigma_1^2 \,\sin^2 \beta
\,\,\,\,\,\,\,\,\,\, ;
\,\,\,\,\,\,\,\,\,\,
\sigma''_1 = \frac{\sigma_1 \sigma'_3}{\sigma''_3 }
\end{equation}
\begin{equation}\label{B}
B = \sin \beta \cos \beta
\left ( \frac{ \sigma_1^{2}-{\sigma'_3}^{2} }{ {\sigma''_3}^{2} } \right ).
\end{equation}
Using the relations (\ref{lambdaplus}) and (\ref{lambdatimes}), the definitions
(\ref{Ralpha},\ref{Rbeta}) and defining $b_1$, $b_2$, $b_3$, $b_4$, $b_5$ and $b_6$
as
 	\begin{equation}\label{defb} 
\begin{tabular}{ll}
$b_1= \sin \alpha $ & $b_2= \cos \alpha + A\,\sin  \alpha \, \sin  \beta$
\\
$b_5= -\cos \alpha $ & $b_3= \sin \alpha - A\,\cos  \alpha \, \sin  \beta$
\\
 $b_4=A\, \cos \beta$ & $b_6 = B$.  \end{tabular} 
	\end{equation}
we obtain for the following probability density
 	\[ 
\begin{tabular}{ll}
$dF_3$ & $=g(y'_2;A\,y'_3,\sigma'_2) \, g(y''_1;B\,x_3,\sigma''_1)
\,dy'_2 dy''_1$ \\
\\
 & $=g(b_2 \,x_2;b_3 \,x_1+ b_4 \,x_3 ,\sigma'_2)\,d|b_2 x_2|$ \\
 & $ \, \times g(b_1 \,x_2;b_5 \,x_1 + b_6\, x_3,\sigma''_1)\, d|b_5 x_1|$  \\
\\
 & $=g(b_1 b_2 \,x_2;b_1 b_3 \,x_1+ b_1 b_4\,x_3 ,|b_1| \sigma'_2)\,
d|b_1 b_2 x_2|$ \\
 & $ \, \times g(b_2 b_1 \,x_2;b_2 b_5 \,x_1 + b_2 b_6 \,x_3,|b_2| \sigma''_1)\,
 d|b_2 b_5x_1|$.  
 \end{tabular} 
	\]
Using  (\ref{2Gaussian}) and since $b_1 b_3 - b_2 b_5 = 1$
the spatial distribution of Eq. (\ref{dPS})
can be rewritten in terms of the coordinates  $(x_1,x_2,x_3)$ 
 	\begin{equation}\label{dPSx} 
\begin{tabular}{ll}
$dP_{\rm S}$ & $=g(x_3;0,\sigma_{\rm I}) \, g(x_1;a_{13}\,x_3,\sigma_{\rm II})$
\\ \\
 & $\times g(x_2; a_{21}\,x_1+a_{23} \,x_3 ,\sigma'_{\rm III})\,dx_1 x_2 x_3$,
 \end{tabular} 
	\end{equation}
where the parameters $a_{13}$, $a_{21}$, $a_{23}$,
$\sigma_{\rm I}$, $\sigma_{\rm II}$ and $\sigma'_{\rm III}$
are obtained from 
Eqs. (\ref{defb},\ref{sigmaprime32},\ref{sigmaprimeprime31},\ref{A},\ref{B})
\begin{equation}\label{defsigI}
\sigma_{\rm II}^2 = b_1^2\,{\sigma'_2}^2 + b_2^2 \,{\sigma''_1}^2
\,\,\,\,\,\,\,\,\,\, ;
\,\,\,\,\,\,\,\,\,\,
a_{13} = b_2\,b_6 - b_1\, b_4 
\end{equation}
{\begin{equation}\label{defsigII}
{\sigma_{\rm I}} = {\sigma''_3} 
\,\,\,\,\,\,\,\,\,\, ;
\,\,\,\,\,\,\,\,\,\,
a_{21} = 
\frac{1}{{\sigma_{\rm II}}^2}\,(b_3b_2{\sigma''_1}^2 - b_5\, b_1 {\sigma'_2}^2)
\end{equation}
\begin{equation}\label{defsigprimeIII}
{\sigma'_{\rm III}} = \frac{{\sigma''_1}{\sigma'_2}}{{\sigma_{\rm II}}}
\,\,\,\,\,\,\,\,\,\, ; \,\,\,\,\,\,\,\,\,\,
a_{23} = 
\frac{1}{{\sigma_{\rm II}}^2}\,(b_4b_2{\sigma''_1}^2 - b_6\, b_1 {\sigma'_2}^2)
\end{equation}

The problem of extracting the values of parameters 
($\alpha$,$\beta$,$\gamma$,$\sigma_1$,$\sigma_2$,$\sigma_3$) 
from a given set of observed parameters
($a_{13}$,$a_{21}$,$a_{23}$,$\sigma_{\rm I}$,$\sigma_{\rm II}$,$\sigma_{\rm
III}$) is not straightforward in general. The special case of
a 2-dimensional 
spatial distribution deserves
mentioning however. If one of the dispersions in Eq. 
(\ref{dPS}) vanishes (e.g. $\sigma_1=0$), the previous calculations
lead to
\[
{\sigma''_1}=0\,\,\,\,\,\,\,\,\,\, ; \,\,\,\,\,\,\,\,\,\,
{\sigma'_2}^2 = \frac{{\sigma_{\rm II}}^2}{\sin^2 \alpha}
\,\,\,\,\,\,\,\,\,\, ; \,\,\,\,\,\,\,\,\,\,
{\sigma'_3}^2 = \frac{{\sigma_{\rm I}}^2}{\cos^2 \beta}
\]
\[
A = -\frac{{\cos \beta}}{\sin \alpha}\,(a_{13}+\cos \alpha
\,\tan \beta)
\,\,\,\,\,\,\,\,\,\, ; \,\,\,\,\,\,\,\,\,\,
{B}=- \tan \beta, 
\]
which implies 
\begin{equation}\label{alphabeta}
{\tan \alpha} = -\frac{1}{{a_{21}}}
\,\,\,\,\,\,\,\,\,\, ; \,\,\,\,\,\,\,\,\,\,
{\tan \beta} = -a_{23} \,\sin \alpha 
\end{equation}
and finally yields
\begin{equation}\label{gamma}
\sinh (\ln ({\tan \gamma})) = -\frac{1}{A}
\,\left ( A^2 - 1 +\frac{{\sigma_{\rm II}}^2}{{\sigma_{\rm I}}^2}
\,\frac{\cos^2 \beta}{\sin^2 \alpha} \right )
\end{equation}
\begin{equation}\label{sigma2}
\sigma_2^2 = \left (1+\frac{A}{\tan \gamma} \right )\,
\frac{{\sigma_{\rm I}}^2}{\cos^2 \beta}
\end{equation}
\begin{equation}\label{sigma3}
\sigma_3^2 = \left (1 - A\,{\tan \gamma} \right )\,
\frac{{\sigma_{\rm I}}^2}{\cos^2 \beta}.
\end{equation}

\section{Distribution in the phase space}
\label{AppendixA3}

The phase space distribution
of the considered dynamical system can be split as
follows
\begin{equation}\label{dPphasespace}
dP_{\rm PS}=dP_{\rm K} \times dP_{\rm S},
\end{equation}
where $dP_{\rm S}$ is the 3D spatial distribution of
Eq. (\ref{dPS}) and the probability density $dP_{\rm K}$ 
describes the distribution of peculiar velocities 
${\bf v}=(v_1,v_2,v_3)$. We chose the system of coordinates
such that the $x_2$-axis points
toward the mean angular position of the cluster
and such that the $x_1$-axis is parallel to the galactic plane;
$v_1$, $v_2$ and $v_3$ are thus components
of the galaxies peculiar velocity respectively along $x_1$, $x_2$
and $x_3$ axes. 
In this frame the 3D velocity
distribution is given by the 3D mean peculiar velocity of the cluster
plus random isotropic components of velocity dispersion
$\sigma_v$, i.e.
\begin{equation}\label{dPK}
dP_{\rm K}=g(v_1;V_1,\sigma_v)dv_1 \,
g(v_2;V_r,\sigma_v)dv_2 \,
g(v_3;V_3,\sigma_v)dv_3,
\end{equation}
where $V_r$ is the radial component of the bulk flow
and $V_1$ and $V_3$ are its components perpendicular to
the line-of-sight.
The distance of the cluster is $D$
and the angular positions with respect to the center
of the cluster are $\theta_1$ and $\theta_3$. In
the approximation of small angles, $\theta_1$ and $\theta_3$
are
\begin{equation}\label{thetha1theta3}
\theta_1 \approx \frac{x_1}{D} 
\,\,\,\,\,\,\,\,\,\, ; \,\,\,\,\,\,\,\,\,\,
\theta_3 \approx \frac{x_3}{D}. 
\end{equation}
Then the redshift $z$ of a galaxy expressed in
km s$^{-1}$ units  is
 	\begin{equation}\label{redshift} 
\begin{tabular}{ll}
$z$ & $=H_0\,D + {\bf v.u}$
\\ 
 & $\approx 
H_0\,D + v_2 + H_0\,x_2 + \theta_1 \,v_1 + \theta_3 \,v_3 $, 
 \end{tabular} 
	\end{equation}
where $H_0$ is the Hubble's constant and ${\bf u}$
is the line-of-sight velocity.

Integrating the probability density $dP_{\rm K}$
over the two unobserved tangential components
of the velocity $v_1$ and $v_3$ we have  
\begin{equation}\label{dPz}
dP_z=g(z;H_0\,D + V_r + H_0 \,x_2  + \theta_1 \,v_1 
+ \theta_3 \,v_3,\sigma_v)\,dz
\end{equation}
Where (\ref{lambdaplus}), (\ref{lambdatimes}) and
(\ref{2Gaussian}) have been used.
It follows from this result and from the expression Eq. (\ref{dPSx}) for $dP_{\rm S}$
the integral $dP_{\rm PS}$ in Eq.
(\ref{dPphasespace})  over the unobserved variable $x_2$ yields
 	\begin{equation}\label{dPobs} 
\begin{tabular}{ll}
$dP_{\rm obs}$ & $=g(z;H_0\,D + V_r + A_{21}\,\theta_1 
+ A_{23} \,\theta_3 ,\sigma_{\rm obs}) $
\\ \\
 & $\, \times g(\theta_1; A_{13}\,\theta_3 ,\sigma'_{\theta_1})
\, g(\theta_3; 0,\sigma_{\theta_3})\,d\theta_3 d\theta_1 dz,
$ 
 \end{tabular} 
	\end{equation}
where the parameters $A_{13}$, $A_{21}$, $A_{23}$,
$\sigma_{\rm obs}$, $\sigma_{\theta_3}$ and $\sigma'_{\theta_1}$
are related to 
$a_{13}$, $a_{21}$, $a_{23}$,
$\sigma_{\rm I}$, $\sigma_{\rm II}$ and $\sigma'_{\rm III}$
from Eqs.  (\ref{defsigI},\ref{defsigII},\ref{defsigprimeIII}).
This characterize the spatial structure of the
cluster, and velocity dispersion $\sigma_v$
and cluster's tangential velocities $V_1$ and $V_3$ 
through the following formulae
\begin{equation}\label{defsigobs}
\sigma_{\rm obs}^2 = {\sigma_v}^2 + {\sigma'_{\rm III}}^2
\,\,\,\,\,\,\,\,\,\, ;
\,\,\,\,\,\,\,\,\,\,
A_{13} = a_{13}
\end{equation}
\begin{equation}\label{A21A23}
A_{21} = V_1 + a_{21}\,H_0 \,D
\,\,\,\,\,\,\,\,\,\, ;
\,\,\,\,\,\,\,\,\,\,
A_{23} = V_3 + a_{23}\,H_0 \,D
\end{equation}
{\begin{equation}\label{defsigtheta1theta3}
\sigma'_{\theta_1} = \frac{{\sigma_{\rm II}}}{H_0 \,D}
\,\,\,\,\,\,\,\,\,\, ;
\,\,\,\,\,\,\,\,\,\,
\sigma_{\theta_3} = \frac{{\sigma_{\rm I}}}{H_0 \,D}.
\end{equation}
So far as the angles $\theta_1$ and $\theta_3$ have been
defined in a way that their averages over the sample
vanish (i.e. $\langle \theta_1 \rangle=0$ and
$\langle \theta_3 \rangle=0$), the observed probability density
$dP_{\rm obs}$ of Eq. (\ref{dPobs}) can be rewritten as
 	\begin{equation}\label{dPobs1} 
\begin{tabular}{ll}
$dP_{\rm obs}$ & $=g(Z; A_{21}\,\theta_1 
+ A_{23} \,\theta_3 ,\sigma_{\rm obs}), $
\\ \\
 & $\, \times g(\theta_1; A_{13}\,\theta_3 ,\sigma'_{\theta_1})
\, g(\theta_3; 0,\sigma_{\theta_3})\,d\theta_3 d\theta_1 dZ,
$ 
 \end{tabular} 
	\end{equation}
where the observable $Z$ is defined as
\begin{equation}
\label{grandZ}
Z=z -\langle z \rangle = z - (H_0 \,D + V_r).
\end{equation}
From Eq. (\ref{dPobs1}) it turns out that the parameters
$A_{13}$, $A_{21}$ and $A_{23}$ can be obtained
using  a standard multiple regression technique, i.e.
{\begin{equation}\label{regressionA13}
A_{13} = \frac{{\rm Cov}(\theta_1,\theta_3)}
{{\rm Cov}(\theta_3,\theta_3)},
\end{equation}
{\begin{equation}\label{regressionA23}
A_{23} = \frac{
{\rm Cov}(\theta_1,\theta_3){\rm Cov}(\theta_1,Z)
-
{\rm Cov}(\theta_1,\theta_1){\rm Cov}(\theta_3,Z)
}
{
{\rm Cov}(\theta_1,\theta_3){\rm Cov}(\theta_1,\theta_3)
-
{\rm Cov}(\theta_1,\theta_1){\rm Cov}(\theta_3,\theta_3),
}
\end{equation}
{\begin{equation}\label{regressionA21}
A_{21} = \frac{
{\rm Cov}(\theta_3,\theta_3){\rm Cov}(\theta_1,Z)
-
{\rm Cov}(\theta_1,\theta_3){\rm Cov}(\theta_3,Z)
}
{
{\rm Cov}(\theta_1,\theta_1){\rm Cov}(\theta_3,\theta_3)
-
{\rm Cov}(\theta_1,\theta_3){\rm Cov}(\theta_1,\theta_3),
}
\end{equation}
where ${\rm Cov}(x,y)=\langle (x-\langle x \rangle)
(y- \langle y \rangle) \rangle$ is the covariance of
random variables $x$ and $y$. Estimates of the dispersions
$\sigma_{\rm obs}$, $\sigma_{\theta_3}$ and $\sigma'_{\theta_1}$
are obtained using
{\begin{equation}\label{sigmatheta3}
{\sigma_{\theta_3}}^2 = {\rm Cov}(\theta_3,\theta_3),
\end{equation}
{\begin{equation}\label{sigmaprimetheta1}
{\sigma'_{\theta_1}}^2 = {\sigma_{\theta_1}}^2
-2\,A_{13}\,{\rm Cov}(\theta_1,\theta_3)
+ A_{13}^2\,{\sigma_{\theta_3}}^2,
\end{equation}
{\begin{equation}\label{sigmaobs}
{\sigma_{\rm obs}}^2 = {\sigma_{Z}}^2
-2\,A_{21}\,{\rm Cov}(\theta_1,Z)
-2\,A_{23}\,{\rm Cov}(\theta_3,Z),
\end{equation}
\[
\,\,\,\,\,\,\,\,\,\,
\,\,\,\,\,\,\,
+ A_{21}^2\,{\sigma_{\theta_1}}^2
+ A_{23}^2\,{\sigma_{\theta_3}}^2
-2\,A_{21} \, A_{23}\,{\rm Cov}(\theta_3,\theta_1)
\],
where $\sigma_{\theta_1}$ and $\sigma_{Z}$ are defined as
{\begin{equation}\label{sigmatheta1Z}
{\sigma_{\theta_1}}^2 = {\rm Cov}(\theta_1,\theta_1)
\,\,\,\,\,\,\,\,\,\,;
\,\,\,\,\,\,\,\,\,\,
{\sigma_{Z}}^2 = {\rm Cov}(Z,Z).
\end{equation}
Finally, the standard deviations i.e. accuracies of the estimators
of parameters $A_{13}$, $A_{21}$ and $A_{23}$ given 
Eqs. (\ref{regressionA13},\ref{regressionA23},\ref{regressionA21})
can be represented as a function of the number $N$ of galaxies in the cluster
{\begin{equation}\label{DeltaA13}
\Delta A_{13} = \frac{1}{\sqrt{N}}
\,\,\frac{\sigma'_{\theta_1}}{\sigma_{\theta_3}},
\end{equation}
{\begin{equation}\label{DeltaA23}
\Delta A_{23} = \frac{1}{\sqrt{N}}
\,\,\frac{1}{\sqrt{1-\rho_{31}^2}}
\,\,\frac{\sigma_{\rm obs}}{\sigma_{\theta_3}},
\end{equation}
{\begin{equation}\label{DeltaA21}
\Delta A_{21} = \frac{1}{\sqrt{N}}
\,\,\frac{1}{\sqrt{1-\rho_{31}^2}}
\,\,\frac{\sigma_{\rm obs}}{\sigma_{\theta_1}},
\end{equation}
where $\rho_{31}$ is the correlation coefficient between
variables $\theta_1$  and $\theta_3$, i.e.
{\begin{equation}\label{rho13}
\rho_{31}^2 = \frac{
{\rm Cov}(\theta_1,\theta_3)\,{\rm Cov}(\theta_1,\theta_3)
}.
{
{\rm Cov}(\theta_1,\theta_1)\,{\rm Cov}(\theta_3,\theta_3)
}
=
A_{13}^2\,\frac{\sigma_{\theta_3}^2}{\sigma_{\theta_1}^2}
\end{equation}

\newpage

Figure captions.
\vspace{0.2in}

Figure 1. 

Virgo subgroup I: 54 galaxies. The top left panel visualizes 
the 3D information contained in the data. The $\theta_1 \theta_3$
plane is perpendicular to the mean line-of-sight 
($\langle l \rangle=284.162$,$\langle b \rangle=73.658$). The
redshift  depth is proportional to the radius of the circles,
backside galaxies are shadow, frontside plain. The 3 other
plots are the projection of this 3D distribution on plane
$x_1x_3$, $x_2x_3$ and $x_1x_2$ respectively 
where $x_2=z-\langle z \rangle$.

\vspace{0.2in}

Figure 2.

Virgo subgroup I: Variation in function of
$\sigma'_{\rm III}$ of the parameters
characterizing the ellipsoidal spatial structure ($\sigma_1$,
$\sigma_2$,$\sigma_3$) 
and its orientation ($\alpha$,$\beta$,$\gamma$). 
The dispersion $\sigma'_{\rm III}$ reads in terms of 
$\sigma_{\rm obs}=167.22$ km s$^{-1}$ and the velocity dispersion
$\sigma_v$ as 
$\sigma'_{\rm III}=\sqrt{\sigma_{\rm obs}^2-\sigma_v^2}$.

\vspace{0.2in}

Figure 3.

Same caption as fig. 2 but for Virgo
subgroup II: The observed dispersion
is $\sigma_{\rm obs}=171.26$ km s$^{-1}$.


\begin{thebibliography}{999}

\bibitem{GR97}
Gurzadyan V.G., Rauzy S., Astrofizika 40, 473 (1997).

\bibitem{Gi}
Girardi M., et al, Astrophys.J. 482, 41 (1997).

\bibitem{GurM}
Gurzadyan V.G., Mazure A., Observatory 116, 391 (1996).

\bibitem{GurM97}
Gurzadyan V.G., Mazure A., MNRAS 295, 177 (1998).
  
\bibitem{GurM01}
Gurzadyan V.G., Mazure A., New Astronomy, March (2001).
 
\bibitem{Fed}
Federspiel M., Tammann G.A., Sandage A., Astrophys.J. 495, 115 (1998).

\bibitem{West}
West M., Blakeslee J.P. astro-ph/0008470

\bibitem{Petr}
Petrosian A.R., Gurzadyan V.G., Hendry M., Nikogossian E., Astrofizika 41, 51 (1998).

\bibitem{GR98}
Rauzy S., Gurzadyan V.G., MNRAS 298, 114 (1998).

\bibitem{Yo}
Yosuda N., Fukugita M., Okamura S., Astrophys.J.Supl. 108, 417 (1997).

\bibitem{GF}
Gonzalez A.H., Faber S.M., Astrophys.J. 485, 80 (1997).

\bibitem{Bo}
Bohrniger H., Neumann D.M., Schindler S., Huchra J.P.,  
Astrophys.J. 485, 439 (1997).

\bibitem{Wu}
Wu Xiang-Ping, Xue Yan-Jie, astro-ph/0106355

\bibitem{Hu}
Huchra J.P. et al, CfA Redshift Catalogue, Harvard-Smithsonian Center
for Astrophysics, Cambridge, (1993).

\bibitem{Gur94a}
Gurzadyan V.G., Kocharyan A.A., 
Paradigms of the Large-Scale Universe,
Gordon and Breach (1994).

\bibitem{Gur94b}
Gurzadyan V.G., Harutyunyan V.V., Kocharyan A.A., Astron.Astrophys.
281, 964 (1994).

\bibitem{Amb36}
Ambartzumian V.A., MNRAS 96, 172 (1936).

\end{thebibliography}
\end{document}